\documentstyle[12pt]{article}
         \makeatletter
         \def\thefigure{\@arabic\c@figure}\def\fps@figure{tbp}
         \def\ftype@figure{1}\def\ext@figure{lof}
         \def\fnum@figure{\protect\footnotesize Fig.\ \thefigure}
         \def\thetable{\@arabic\c@table}
         \def\fps@table{tbp}\def\ftype@table{2}\def\ext@table{lot}
         \def\fnum@table{\protect\footnotesize Table \thetable}
         \makeatother
\begin{document}
\vspace*{0.3in}
\begin{center}
  {\Large \bf 
Nuclear Astrophysics in Storage Rings
}\\
  \bigskip
  \bigskip
  {\Large C.A. Bertulani\\
Instituto de F\'\i sica, Universidade
        Federal do Rio de Janeiro\\ 
        21945-970, Rio de Janeiro, RJ, Brazil}%
        \thanks{E-mail: bertu@if.ufrj.br. Work partially supported by the
                Conselho Nacional de Desenvolvimento Cient\'\i fico e
                Tecnol\'ogico, CNPq/Brazil and the Gesellschaft 
                f\"ur Scwerionenforschung, GSI/Germany.}
  \bigskip

   \end{center}
\begin{abstract}
Nuclear reaction cross sections
are usually very small in typical astrophysical
environments. It has been one of the major challenges of experimental
nuclear astrophysics to assess the magnitude of these cross sections in
the laboratory. For a successful experiment
high luminosity beams are needed. Increasing the target width, one also 
increases the reaction yields. But, this is of limited use due to
multiple scattering in the target. Storage rings are a very good way
to overcome these difficulties. In principle, they can be tuned to 
large luminosities, and have the advantage of crossing the
interaction region many times per second (typically 
one million/s), compensating low density internal gas targets,
or low reaction rates in beam-beam collisions. Storage rings are also
ideal tools for precise  measurements of masses and beta-decay lifetimes
of nuclei of relevance for astrophysics. 
\end{abstract}
\baselineskip 4ex

\section{CENTRAL PROBLEMS IN NUCLEAR ASTROPHYSICS}

\subsection{Screening}
 
Nuclear astrophysics requires the knowledge of the reaction rate $R_{ij}$
between the nuclei species $i$ and $j$. It is given by $R_{ij}=n_in_j<\sigma
v>/(1+\delta_{ij})$, where $\sigma$ is the cross section, 
$v$ is the relative velocity between the reaction partners, 
$n_i$ is the number density of the
nuclide $i$, and $<>$ stands for energy average. 
Extrapolation procedures are often needed to obtain cross sections at 
the energy region of astrophysical 
relevance.
While non-resonant cross sections
can be rather well extrapolated to the low-energy region, the
presence of continuum, or subthreshold resonances, complicate these
extrapolations. 
Another problem is that charged-particle
induced reactions are electron-screened in stellar environments
\cite{Sal54}. Applying the Debye-H\"uckel approach, one finds that the plasma
enhances reaction rates, e.g., $^3He(^3He, \ 2p)^4He$ and
$^7Be(p, \ \gamma )^8B$, by as much as 20\%. 
Laboratory nuclear reactions are also
modified by screening effects due to electrons which are 
inevitably present in the
target \cite{Ass87}. As can be seen in figure 1 for the reaction $^3He(d, \ p) ^4He$ 
(probably the
best studied example of laboratory electron screening effects, 
both experimentally and
theoretically - Data are from ref. \cite{Eng88}. Calculations are
from ref. \cite{Blu90}), the effects of laboratory electron screening are 
far from 
being understood and require further investigation. 

While corrections due to plasma screening effects have to rely on theoretical models, 
laboratory screening can be avoided experimentally  using
crossed beams in storage rings. This
might be of great usefulness for the study of the reactions 
$^6Li(d, \alpha)^4He$,$^7Li(p, \alpha)^4He$,\-$^{11}B(p, \alpha)2\ ^4He$, and
$^6Li(p, \alpha)^3He$, where discrepancies between experiment and theoretical
calculations of electron screening are large \cite{Rol95}.

\begin{figure}[htb]
\begin{minipage}[t]{80mm}
{\rule[-20mm]{0mm}{52mm}}
\caption{
The $^3He(d,\ p)^4He$ S-factors. 
Dashed line is the calculated 
astrophysical S-factor for bare nuclei. 
Solid and dashed curves include corrections of screening.
}
\end{minipage}
\hspace{\fill}
\begin{minipage}[t]{75mm}
{\rule[-26mm]{0mm}{72mm}}
\caption{Astrophysical S-factor for $^4He(^3H, \ \gamma)^7Li$ compared to
        calculations (dashed and solid curves).
}
\end{minipage}
\end{figure}

\subsection{Homogeneous Big Bang} 
Deuterons were formed by fusion of neutrons and protons 
when the Universe temperature reached $T=10^{12}$ K. However, 
they were promptly dissociated by  photons which existed in thermal 
equilibrium. Only when $T \sim 10^9$ K deuterons were formed in 
significant concentration and the primordial nucleosynthesis 
bottleneck was overcome. Since, (i) the binding energies of 
$^3He$ and $^4He$ are larger than that of the deuteron and (ii)
the cross sections for capture of protons and neutrons by deuterons
are large, deuteron  was rapidly consumed. 
Standard primordial nucleosynthesis assumes a homogeneous proton-to-neutron 
density. Thus the nuclear reaction flow along the $Z\sim N$ stability line 
stops at $^8B$ because of its instability. Since the element with A=5 does 
not exist, $^4He$ is by far the most abundant product of primordial 
nucleosynthesis. Only rather small traces of $D$, $^3He$ and $^7Li$ are made. 
Just because of that, the relative abundance of these elements are a good 
test of the Big Bang scenario. 
Among the several reactions relevant for primordial nucleosynthesis, 
the radiative capture reaction $^4He(^3H, \ \gamma)^7Li$ deserves special 
attention. 
There has been a large discrepancy among five measured  
cross sections for this reaction, as shown in figure 2 
(adapted from \cite{Kaj95}). This leads to 
30\% - 40\% in the primordial $^7Li$ abundance and also in the determination of
the cosmological parameter $\Omega_B$ \cite{Kaj95}.      

\subsection{Inhomogeneous Big Bang}

Inhomogeneous 
Big Bang models allow for fluctuations in baryon densities that are 
related to different neutron mass fractions at the beginning of 
nucleosynthesis. These  models involve some nuclear reactions of which 
cross sections were not known. 
The postulated dominant flow path, which bypasses the 
A=8 gap, is $^4He(t,\ \gamma)^7Li(n, \ \gamma)\break ^8Li(\alpha, \ n)^{11}B(n, \ 
\gamma)^{12}B(\beta^-\nu)^{12}C$. Here, the $ ^8Li(\alpha, \ n)^{11}B$ 
reaction was completely unknown. The first experiment was made using the 
reverse reaction $^{11}B(n, \ \alpha)^8Li$, finding a large resonance at 
around 0.5 MeV above the $\alpha$ threshold in $^{12}B$ \cite{Par90}. This energy 
corresponds to  the Gamow peak of $10^9$ K, which is a typical temperature 
for heavy elemental synthesis in the primordial site. However, this reaction 
has been measured directly using a low-energy $^8Li$ beam at $E_{c.m.}\ge 
1.5$ MeV \cite{Boy92} at RIKEN. The cross section estimated by an 
extrapolation to $ E_{c.m.} \sim 0.5$ MeV is larger than the previous 
value \cite{Par90} by roughly a factor of 5. The reason for the 
difference between the two results is that the inverse process uses a stable 
$^{11}B$ beam. Thus, only the cross section for the transition to the ground 
state was determined by using the principle of detailed balance. However, 
there are at least four more excited bound states of $^{11}B$ which 
contribute to the production of $^{11}B$ through  $ ^8Li(\alpha, \ 
n)^{11}B^*(\gamma)^{11}B$.  This is a very interesting subject to be 
pursued by a direct measurement at $ E_{c.m.} \sim 0.5$ and shows the 
importance of radioactive beam facilities for astrophysical purposes.

\subsection{The Sun}

Because of the work of Bahcall and collaborators, there seems to be a 
very consistent understanding of the hydrodynamics and the nuclear reaction 
network in the Sun \cite{Bah89}. 
The main  network for energy production in the 
sun is the well known ppI-chain, with the net result of fusing four 
protons to form one tightly bound $^4He$ nucleus. In a 14\% branch, 
a $^7Be$ nucleus is produced via the $^3He+^4He \rightarrow ^7Be+\gamma$ 
radiative capture process.  In a very rare event, 
the $^7Be$ nucleus will capture a proton to form $^8B$, which subsequently 
decays via positron emission to an excited state of $^8Be$; this state, 
in turn, rapidly breaks into two $\alpha$-particles. This is the ppIII chain:
\begin{equation}
p(p, \ e^+\nu)d(p,\ \gamma )^3He(^4He, \ \gamma)^7Be(p, \ 
\gamma)^8B(e^+\nu)^8Be^* \rightarrow 2 \ ^4He \ .
\end{equation}
Although this chain delivers a very small part of the 
energy produced in the sun, it is responsible for the emission of 
high-energy neutrinos.
These high energy neutrinos play an important role in the history of the 
search for solar neutrinos. Terrestrial detection of solar neutrinos have 
lead to the conclusion that an appreciable part of these neutrinos are 
missing. This is known as the {\it solar neutrino problem}, which persists for 
about two decades \cite{Bah89}.
Due to the Coulomb barrier, the reaction cross section $^7Be(p, 
\gamma)^8B$ is very small at the Gamow peak ($\sim 20$ keV). Direct 
measurements have reached $E_{p-Be} \sim 140$ keV. However, discrepancies 
among experimental data are evident as shown
in figure 3. Long ago, Barker \cite{Bar80} has emphasized that 
an analysis of the existing experimental data yields an S-factor for this 
reaction at low energies which is uncertain by as much as 30\%. This 
situation has not changed. A measurement of this reaction at the Gamow peak 
is highly desirable. 

\subsection{Massive Stars}

\subsubsection{The $^{12}C(\alpha, \ \gamma)^{16}O$ reaction}

We now consider a massive star,
of order of 25 solar masses. After hydrogen burning, when
the temperature and the density of the core of the star reaches values of 
$T\sim 1.5 \times 10^8$ K and $\rho \sim 5 \times 10^4 \ g/cm^3$, helium can
react with helium to form $^{12}C$. This 
happens via an $s$-wave resonance ($E=92\ keV$) in $^8Be$, with mean lifetime
close to $10^{-16} \ s$. This lifetime is long enough that a small amount
of $^8Be$ is formed and capture an $\alpha$-particle to form $^{12}C$. The
cross section for this reaction is enhanced by the existence of a resonance
at 287 keV in $^{12}C$ above the $\alpha+^8Be$ threshold. This was predicted 
by Hoyle before the state was confirmed experimentally \cite{Hoy53}. The $^{12}C$ 
produced can capture an $\alpha$-particle to form $^{16}O$. This 
$^{12}C(\alpha, \ \gamma ) ^{16}O$ reaction is a key process for the elemental
abundances of heavy elements, and for the evolution of the stars. It is
argued that the cross section for this reaction should be known to better than
20\%, for a good modeling of the stars \cite{Woo85}. This goal has not yet been 
achieved. 

The problem with the experimental  determination of the 
$^{12}C(\alpha, \ \gamma ) ^{16}O$ reaction at the energy $E \sim 300\
keV$ ($T\sim 2\times 10^8 \ K$) is the existence of a broad $1^-$ 
resonance at $E=2.42 \ MeV$ above the $\alpha+^{12}C$ threshold, as well as
two sub-threshold states: a $1^-$ state just 45 keV below the threshold, and
a $2^+$ state 245 keV below threshold. These states  interfere and their
contribution to the data cannot be separated unambiguously. 
Theoretical calculations of this reaction are also not reliable, as this
process is, in first order isospin forbidden and proceeds via unknown
$T=1$ impurities in the $^{16}O$ compound nuclear states.
Recently, the $E1$ radiative capture contribution to this reaction was 
obtained by the beta-delayed emission of $\alpha$-particles from 
a $2^-$ state in 
$^{16}N(t_{1/2}=7.1\ s)$
\cite{Buch93}. Due to the beta-decay selection rules, excited
states in $^{16}O$ with $J^\pi = 1^-, \ 2^-$, and $3^-$ are populated.
Thus, only $E1$ transitions in the alpha decay are possible. It is expected
however, that $E2$ transitions in the radiative capture reaction
$^{12}C(\alpha, \ \gamma ) ^{16}O$ are as well important as $E1$ transitions.
The total capture cross section, $\sigma_{E1}+\sigma_{E2}$ is therefore 
still unknown with the required precision.
 
\subsubsection{Rapid capture processes}

With increasing temperatures and densities of the core the star reaches
advanced burning stages. Then carbon burning ($T\sim 5\times 10^8 \ K$,
$\rho=1-2 \times 10^5\ g/cm^3$) is followed by neon burning
($T\sim 5\times 10^8 \ K$, $\rho=4\times 10^6\ g/cm^3$), oxygen burning
($T\sim 1.5\times 10^9 \ K$, $\rho=10^7\ g/cm^3$)
and silicon burning cycles
($T\sim 3\times 10^9 \ K$, $\rho=3 \times 10^7\ g/cm^3$). The star then forms 
$^{56}Fe$ which is the end point of spontaneous nuclear fusion. The energy
released by nuclear fusion stops and the core collapses due to gravitation
pressure. The iron nuclei are broken up, absorbing energy from the plasma,
and accelerating the collapse. The high density of the core favors electron 
capture, transforming protons in neutrons and emitting neutrinos, which
escape from the star. When the core density reaches densities of about 
$2.4 \times 10^{14}\ g/cm^3$ (nuclear matter density), it cannot contract
further, bounces back,
and a shockwave will run through the outer layers of the collapsed star.
However, this shockwave does not carry enough energy to explode the star.
The shock is heating the material to such high temperatures that the
previously produced iron is photo disintegrated again. This process takes
4--7\,MeV per nucleon (7\,MeV for a complete photo disintegration into nucleons)
and will eventually halt the shockfront.
The formation of the neutron star leads to a gain in gravitational binding
energy which is released in the form of
neutrinos. Although the interaction of neutrinos with matter is quite
weak, considerable amounts of energy can reheat the outer layers even
if only 1\% of the 10$^{53}$\,erg (10$^{46}$\,J) in neutrinos
is deposited via neutrino captures on neutrons and protons.
This accelerates the shockwave
again and can finally explode the star. Because of the heating and
expansion of the gas, a zone with low density and high temperature will
be formed behind the shockfront, the so--called {\em high--entropy bubble}.

During the subsequent cooling of
the plasma the nucleons will recombine again, first to $\alpha$--particles,
then to heavier nuclei, starting with the reactions
3$\alpha \rightarrow ^{12}$C and $\alpha+\alpha+{\rm n} \rightarrow ^9$Be, followed
by $^9$Be($\alpha$,n)$^{12}$C.
Temperature and density are
dropping quickly in the adiabatically expanding high--entropy bubble.
This will hinder the recombination of alpha particles into heavy nuclei,
leaving
a high $Y_n/Y_{\rm seed}$ and sufficient neutrons for an r--process,
(acting on the newly produced material)
at the end of the $\alpha$--process after freeze--out of charged particle reactions.
Approximately half of all stable nuclei observed in nature in the heavy
element region about $A>60$ is produced in the r--process. 
This r--process occurs
in environments with large neutron densities which lead to
$\tau_{\rm n} \ll \tau_{\beta}$.

\begin{figure}[htb]
\begin{minipage}[t]{80mm}
{\rule[-26mm]{0mm}{72mm}}
\caption{S-factors for $^7Be(p, \ ^8Be)$ from several experiments. Curves
are theoretical calculations based on a potential model.}
\end{minipage}
\hspace{\fill}
\begin{minipage}[t]{75mm}
{\rule[-26mm]{0mm}{72mm}}
\caption{Level density (in levels per MeV) at the respective neutron
separation energy. Higher level densities are represented by
darker dots.}
\end{minipage}
\end{figure}

The most neutron--rich isotopes along the r--process
path have lifetimes of less than one second and more typically 10$^{-2}$
to 10$^{-1}$\,s. Cross sections for most of the nuclei involved cannot
be experimentally measured due to the short half--lives. Therefore,
theoretical descriptions of the capture cross sections as well
as the beta--decay half--lives are the only source of the nuclear physics
input for r--process calculations. For nuclei with about $Z>80$ beta--delayed
fission and neutron--induced fission might also become important
(for a review, see \cite{Ober96}).

\section{COSMOCHRONOMETRY}

Presently, the most important application of storage rings to nuclear astrophysics
has been the determination of {\it bound-$\beta$-decay} lifetimes.
An ingenuous experiment performed with the heavy-ion storage
ring ESR at GSI/Darmstadt measured the $\beta$-decay half-life of bare
$^{163}Dy$ \cite{Jun92}. 
The $^{163}Dy^{66+}$ ions were stored and accumulated in the
ring. Contrary to neutral atoms the electron originated from $\beta$-decay of
bare nuclei can occupy one of the atomic orbits in the nucleus. This reduces
dramatically the $\beta$-decay half-lives of several nuclei, or can be the only way
that the $\beta$-decay can proceed. In fact, the
experiment deduced a half-life of $47{+5\atop -4}$ d for bare $^{163}Dy$, 
which is stable as a neutral atom.

More recently \cite{Bos95}, 
the same method was used to study the bound $\beta$-transitions in
bare $^{187}Re$. The observed (preliminary) 
half-life of $34 \pm 7\ yr$  is much smaller than the
neutral atom half-life, $T_{1/2}=43 \ Gyr$ [$1\ Gyr = 10^9 \ yr$]. This result has 
a great influence on the use of the $^{187}Re-^{187}Os$ as a 
cosmochronometer isotopic
pair. The {\it astration} of $^{187}Re$  in the hot environment of next generation stars
may involve bare, or almost bare, $^{187}Re$ nuclei, and induces a big
difference on the
derivation of the Galactic age from the $^{187}Os/^{187}Re$ relative abundance \cite{Bos95}. 

There is a proposal to measure the bound $\beta$-decay of bare $^{205}Tl$
\cite{Bos96}. This
nucleus can be transmuted to $^{205}Pb$ by solar neutrino capture. Thus, a measurement
of the $^{205}Tl/^{205}Pb$ abundance ratio in deep lying rocks can give the information
on the solar neutrino flux integrated over $10^7 \ yr$. The neutrino capture goes to the
first-excited state of $^{205}Pb$ and the experiment aims in providing this
nuclear transition matrix element which is unknown. 

The subject of bound $\beta$-decay experiments is
discussed in more details by O. Keppler in this proceedings.

\section{DETERMINATION OF ASTROPHYSICAL NUCLEAR RATES}

To calculate astrophysical reaction rates with some reliability,
one needs to know several nuclear properties which have to be assessed
experimentally. Here we give some examples of  what can be 
studied with storage rings.

\subsection{Elastic Scattering and ${\bf (p, \ p')}$ reactions}

The use of internal proton gas targets is a standard technique in
storage rings. Protons are a very useful probe since their internal
structure remains unaffected during low energy collisions. Nuclear densities
are a basic input in 
theoretical calculations of astrophysical reactions at low energies.
These can be obtained in, e.g., elastic proton
scattering. Elastic scattering in high energy collisions essentially
measures the Fourier transform of the matter distribution. Considering
for simplicity the one-dimensional case, for light
nuclei one has $\int e^{iqx} \rho(x) dx \sim \int e^{iqx} [a^2+x^2]^{-1}
= (\pi/a).e^{-|q|a}$, where
$q=2k\sin \theta/2$, for a c.m. momentum $k$, and
a scattering angle $\theta$. For heavy nuclei the density 
$\rho$ is better described by Fermi function, and
$\int e^{iqx} [1+e^{(x-R)/a}]^{-1} \sim (4\pi) . \sin qR . e^{-\pi q a}$, for
$R>>a$, and $qa>>1$. Thus, the distance between minima in elastic 
scattering cross sections
measures the nuclear size, while its exponential decay dependence reflects
the surface diffuseness.

During the last few years, elastic proton scattering has been one of the 
major sources of information on the matter distribution of unstable
nuclei in radioactive beam facilities. The extended matter 
distribution of light-halo nuclei ($^8He,\ ^{11}Li, \ ^{11}Be,$ etc.)
was clearly identified in recent elastic scattering experiments
\cite{Neu95,Kor96}.
Information on the matter distribution of many 
nuclei important for the nucleosynthesis in inhomogeneous Big Bang and
in r-processes scenarios could also be obtained in elastic scattering
experiments. Due to the loosely-bound character and
small excitation energies of many of these nuclei, high 
energy resolution is often necessary. These can be achieved in
storage rings by means of electron cooling, or stochastic cooling.        

In $(p, \ p')$ scattering one obtains information on the excited
states of the nuclei. For the same reason as in the elastic scattering
case, high energy resolutions obtained in storage rings are of 
crucial relevance \cite{Kor96}.

\subsection{Transfer Reactions}

The cross section for transfer reactions $A(a,b)B$ are 
given by
\begin{equation}
{d\sigma \over d\Omega} \propto \sum_{lj} S_{AB}(lj) |M_{lj}|^2 
\ ,
\end{equation}
where $S_{AB}$ is the spectroscopic factor for the overlap
integral of the wave functions of nuclei $A$ and $B$. If
$a=b+x$ (stripping reaction), 
this integral is given by $\psi_x({\bf r}_{xB})=
\int d\xi \Phi_A^*\ \Phi_B$. In eq. (2),  $lj$ denotes
the angular momentum quantum numbers of the particle $x$ within
$B$ and $M_{lj}$ is an integral of the product of the interaction 
$V_{bx}$, the incoming and outgoing scattering states $\chi_a^{(+)}$
($\chi_x^{(-)}$), and the wave function $R_{njl}(r_{xB})$. 
Transfer reactions are a well established
tool to obtain spin, parities, energy, and spectroscopic factors
of states in the $B+x$ system. Experimentally, 
$(d, \ p)$ reactions are mostly used due to the simplicity of
the deuteron.

The astrophysically relevant nuclear reactions proceed via 
compound-nucleus reactions (CN) or direct reactions (DR).
The decisive mechanism depends on the number of levels in the CN.
If there are no resonances in the energy interval relevant for
the reaction, one can use the DR models, like direct capture (DC).
In this case, $\sigma^{nr} \propto \sum_x S_x \sigma_x^{DC}$, where
the sum extends over all bound states in the final nuclei and $S_x$
is the relevant spectroscopic factor. For radiative capture reactions, e.g.,
$b(x, \gamma)a$, 
$\sigma_x^{DC}\propto |<\chi_b\chi_x|{\cal O}_{\pi\lambda}|\phi>|^2$, where
$\chi$ are scattering waves, $\phi$ is the bound-state wave function, and
${\cal O}_{\pi\lambda}$ is the electromagnetic operator.

In the case of a single isolated resonance the resonant part of the
cross section is given by the Breit-Wigner formula
$\sigma_r (E) \propto \Gamma_{in} \ \Gamma_{out} \
[(E_r-E)^2+\Gamma_{tot}^2/4]^{-1}$. The partial widths of the entrance 
and exit channels are $\Gamma_{in}$ and $\Gamma_{out}$, respectively.
The total width $\Gamma_{tot}$ is the sum over the partial widths of
all channels. The particle width $\Gamma_p$ can be related to the
spectroscopic factors $S$ and the single-particle width $\Gamma_{s.p.}$
by $\Gamma_p= C^2 S \Gamma_{s.p.}$, where $C$ is the isospin Clebsch-Gordan 
coefficient. 

If one considers a few resonant states the R-matrix theory is appropriate.
In this theory, the resonant part of the
cross sections is obtained by a sum of Breit-Wigner
functions, slightly displaced (Thomas-Lane correction)
in the resonant energy due to the background
terms of the other resonances that are superimposed to the considered
resonance. 

We
see that $(d,\ p)$ reactions are important to astrophysics indirectly to
determine the spectroscopic factors for $(n,\ \gamma)$. 
Many of these are unknown and urgently needed for theoretical
calculations. 
Examples are the reactions $^7Li(n,\ \gamma)^8Li$ 
and $^9Be(n, \ \gamma )^{10}Be$
where spectroscopic factors from the $(d,\ p)$ reactions are not known
experimentally. 
Interesting to astrophysics are  most
$(d,\ p)$ reactions on the neutron rich side of the nuclide card,
where no $(d,\ p)$ has been measured before.

If the level density of the CN is so high that there are many overlapping
resonances in a certain energy interval the statistical Hauser-Feshbach
model can be applied. In an astrophysical plasma
one needs to know the transmission coefficients 
thermally averaged over several populated states  
\begin{eqnarray}
T (E,J,\pi) & = & \sum^{\nu_m}_{\nu =0}
T^\nu(E,J,\pi,E^\nu_m,J^\nu_m, \pi^\nu_m) \\\nonumber
 & & + \int\limits_{E^{\nu_m}_m}^{E-S_{m}} \sum_{J_m,\pi_m}
T\left(E,J,\pi,E_m,J_m,\pi_m\right)
\rho\left(E_m,J_m,\pi_m\right) dE_m  \quad ,
\end{eqnarray}
where $S_m$ is the channel separation energy, and the summation over excited
states above the highest experimentally known state $\nu_m$ is changed 
to an integral over
the level density $\rho$. The energy, spin and parities of the states
are denoted by $E_m$, $J_m$ and $\pi_m$, respectively.

The necessary condition for application of statistical models
is a large number of resonances
at the appropriate bombarding energies, so that the cross section can be
described by an average over resonances. 
In the case of neutron--induced reactions a criterion for the
applicability can directly be derived from the level density. 
The relevant energies will lie very close
to the neutron separation energy. Thus, one only has to consider the
level density at this energy. It is usually said that
there should be at least 10 levels per MeV for reliable statistical
model calculations. The level densities at the appropriate neutron
separation energies are shown in Fig. 4, adapted from
ref. \cite{Ober96} (note that therefore
the level density is plotted at a {\em different} energy for each
nucleus). One can easily identify the magic neutron numbers by the drop
in level density. A general
sharp drop is also found for nuclei close to the neutron drip line. For
nuclei with such low level densities the statistical model cross
sections will become very small and other processes might become
important, such as direct reactions  \cite{Ober96}.
The above plot can give hints on when it is safe to use the statistical
model approach.

The use of $(d,p)$ reactions at storage rings has already 
been proposed for the spectroscopy of neutron resonances in
r-processes nuclei close to the magic shells \cite{Egel96}. 
The two proposed cases (i) $^{46}K(d, \ p)$ and (ii) $^{134}Te(d, \ p)$
are of interest in connection with the interpretation of (i) the 
$^{48}Ca/^{46}Ca$-abundance anomaly observed in the Allende metereorite
\cite{San82} and (ii) the $A\sim 130$ r-process abundance peak. 
For both isotopes a measurement of the spectroscopic factors and 
spins for transitions to the ground state and the lowest
excited states will allow improved calculations of the
DR cross section, which is the expected to be the dominant
component of the neutron capture rate.

Mass and half-life measurements of neutron deficient
isotopes is currently of  strong interest. Recently, 
network calculations have been extended up to the mass 100 (Z=50) 
and
different mass formula  have been tested
(Hilf, Jaenecke, M\"oller, etc.) \cite{Wie96}. For the half lives 
QRPA and partly shell model calculations were used. 
It turns out that the reaction path and also the time scale for the 
rapid proton capture process are strongly determined by the masses. This is in 
particular very sensitive in the mass 80 region where large deformation 
effects have to be considered. 
Also to be measured need to be the decay of the isotopes along the 
N=Z line in this mass range. The measurement of $^{80}Zr$ has just been
completed and of $^{84}Mo$ has been proposed \cite{Wie96}. 
But, there is still a lot more to be done. 
\subsection{Trojan Horse}
In order for transfer reactions $A(a,b)B$ to be effective, a matching condition
between the transferred particle and the internal particle momenta has
to exist. Thus, beam energies should be in the range of a few 10 MeV per
nucleon. It has been proposed  that low energy
reactions of astrophysical interest
can be extracted directly from breakup reactions $A+a \longrightarrow b+c+B$ by
means of the Trojan Horse effect \cite{Bau86}. 
If the Fermi momentum of the particle
$x$ inside $a=(b+x)$ compensates for the initial projectile velocity $v_a$,
the low energy reaction $A+x=B+c$ is induced at very low (even vanishing)
relative energy between $A$ and $x$.
To show this, one writes the DWBA cross section for the breakup reaction as
$d^3/d\Omega_b d\Omega_c dE_b \propto |\sum_{lm} T_{lm}({\bf k_a, k_b, k_c})
S_{lx} Y_{lm}({\bf k_c})|^2$, where $T_{lm}= <\chi_b^{(-)} Y_{lm} f_l
|V_{bx}|\chi_a^{+}\phi_{bx}>$. The threshold behaviour $E_x$ for the
breakup cross section $\sigma_{A+x\rightarrow B+c}
=(\pi/k_x^2)\sum_l (2l+1)|S_{lx}|^2$ is well known: since
$|S_{lx}|\sim \exp(-2\pi \eta)$, then $\sigma_{A+x\rightarrow B+c}\sim
(1/k_x^2)\ \exp(-2\pi \eta)$. In addition to the threshold
behaviour of $S_{lx}$, the breakup cross section is also governed by the
threshold behaviour of $f_l(r)$, which for $r\longrightarrow \infty$
is given by $f_{l_x}\sim (k_xr)^{1/2} \ \exp(\pi \eta ) \
K_{2l+1}(\xi)$, where $K_l$ denotes the Bessel function of the second
kind of imaginary argument. The quantity $\xi$ is independent of $k_x$
and is given by $\xi=(8r/a_B)^{1/2}$, where $a_B=\hbar^2/mZ_AZ_xe^2$ is the
Bohr length. From  this one obtains that  
$d^3/d\Omega_b d\Omega_c dE_b {\ \ \  \longrightarrow \ \ \ \atop 
{E_x \rightarrow 0}} {\rm const.}$. The coincidence cross section tends
to a constant which will in general be different from zero. This is
in striking contrast to the threshold behavior of the two particle reaction
$A+x=B+c$. The strong barrier penetration effect on the
charged particle reaction cross section is canceled completely by the behaviour
of the factor $T_{lm}$ for $\eta \rightarrow \infty$.
Thus, from a measurement of the breakup
reaction   $A+a \longrightarrow b+c+B$ and a theoretical calculation of
the factors $T_{lm}$ one could extract the $S$-matrix elements $S_{lx}$ needed
for the reaction $A+x=B+c$. Basically, this technique extends the method 
of transfer reactions to continuum states.

Among candidates to be studied with this method at storage rings, one
might consider to investigate the reaction $^{18}O(p, \ \alpha )^{15}N$
by means of the reaction $^{18}O+(b+p)\longrightarrow b+\alpha+^{15}N$
where $a=(b+p)=d, \ ^3He, \ \alpha, ...$. The same could be applied to
the reactions $^{15}N(p, \ \alpha)$ or $^{17}O(p,\ \alpha)$, relevant for the
CNO cycle. Also, the reaction $\alpha+^3He \rightarrow ^7Be+\gamma$, which is
relevant for the solar neutrino problem, could be investigated e.g. in the
reaction $^{16}O+^3He \rightarrow ^{12}C+\gamma +^7Be$.

\subsection{Asymptotic Normalization Coefficients}

The amplitude for the radiative capture cross section $b+x\longrightarrow a+ \gamma$
is given by $M=<I_{bx}^a({\bf r_{bx}})|{\cal O}({\bf r_{bx}})|
\psi_i^{(+)}({\bf r_{bx}})>$, where 
$I_{bx}^a=<\phi_a(\xi_b, \ \xi_x,\ {\bf r_{bx}})
|\phi_x(\xi_x)\phi_b(\xi_b)>$ is the integration over the internal coordinates
$\xi_b$, and $\xi_x$, of $b$ and $x$, respectively.
For low energies, the overlap integral $I_{bx}^a$ is dominated by contributions
from large $r_{bx}$. Thus, what matters for the calculation of the matrix element
$M$ is the asymptotic value of   $I_{bx}^a\sim C_{bx}^a \ W_{-\eta_a, 1/2}(2\kappa_{bx}
r_{bx})/r_{bx}$, where $C_{bx}^a$ is the asymptotic normalization coefficient (ANC) and
$W$ is the Whittaker function. This
coefficient is the product of the spectroscopic factor and a normalization constant
which depends on the details of the wave function in the interior part of the
potential. Thus,  $C_{bx}^a$ is the only unknown factor needed to calculate
the direct capture cross section. These normalization coefficients
can be found from: 1) analysis of classical nuclear reactions such as elastic scattering
[by extrapolation of the experimental scattering phase shifts to the bound 
state pole in the energy plane], or 2) 
peripheral transfer reactions whose amplitudes contain the
same overlap function as the amplitude of the corresponding astrophysical radiative
capture cross section \cite{Muk90}.

To illustrate this technique, let us consider the proton transfer
reaction $A(a,b)B$, where $a=b+p$, $B=A+p$.   Using the asymptotic form of the
overlap integral the DWBA cross section is given by
$d\sigma/d\Omega = \sum_{J_Bj_a}[(C_{Ap}^a)^2/\beta^2_{Ap}][(C_{bp}^a)^2/\beta^2_{bp}]
{\tilde \sigma}$ where 
$\tilde \sigma$ is the reduced cross section
not depending on the nuclear structure, 
$\beta_{bp}$ ($\beta_{Ap}$) are the asymptotic normalization of the
shell model bound state proton wave functions in nucleus $a (B)$
which are related to the corresponding ANC's of the overlap function as
$(C_{bp}^a)^2 =S^a_{bp}  \beta^2_{bp}$. Here $S^a_{bp}$ is the spectroscopic factor.
Suppose the reaction $A(a,b)B$ is peripheral. Then each of the bound state
wave functions entering $\tilde \sigma$ can be approximated by its asymptotic form
and $\tilde \sigma \propto \beta_{Ap}^2 \beta_{bp}^2$. Hence  
$d\sigma/d\Omega = \sum_{j_i}(C_{Ap}^a)^2(C_{bp}^a)^2 R_{Ba}$ where
$R_{Ba}={\tilde \sigma}/\beta^2_{Ap} \beta^2_{bp}$ is independent of
$\beta^2_{Ap}$ and $\beta^2_{bp}$. Thus for surface reactions the DWBA cross 
section is actually parameterized in terms of the product of the square of the
ANC's of the initial and the final nuclei $(C_{Ap}^a)^2(C_{bp}^a)^2$ rather
than spectroscopic factors. This effectively removes the sensitivity in the
extracted parameters to the internal structure of the nucleus.

Recently, 
this technique has been applied to the reaction
$^7Be(d,n)$ at $E_{c.m.}=5.8$ MeV
to determine the asymptotic normalization for 8B
\cite{Wei96}. The astrophysical $S_{17}(0)$ factor for the
$^7Be(p, \ \gamma )^8B$ reaction was derived to be 
$27.4 \pm 4.4$ eV.b through the asymptotic normalization
coefficient extracted from the experimental data.
This is a rather high value, compared to previous measurements,
as can be seen from figure 3.

\subsection{Charge-exchange, $(p,n)$, Reactions}
Charge exchange induced in $(p,n)$ reactions are often used to obtain
values of Gamow-Teller matrix elements which cannot be extracted from 
beta-decay experiments. This approach relies
on the similarity in spin-isospin space of
charge-exchange reactions and $\beta$-decay operators. As a result of
this similarity, the cross section $\sigma(p, \ n)$ at small momentum
transfer $q$ is closely proportional to $B(GT)$ for strong transitions
\cite{Tad87}.

The efficiency of many $\nu$ detectors ($^{37}Cl$, the proposed $^{115}I$
and $^{127}I$ detectors, etc.) depends on the cross section for absorption 
of neutrinos, through inverse $\beta$ decay, in the detector material. When
corresponding matrix elements for allowed GT beta decay 
cannot be measured in $\beta$ decay experiments, for example, if the 
$\beta$ decay is not energetically allowed, it is common to obtain these
matrix elements using charge exchange reactions.

Such experiments could be carried out using storage rings. However, there
is some debate if the accuracy of $(p, \ n)$ reactions as a calibration
of solar neutrino detectors can be achieved. As shown in ref. \cite{Aus94}, for important 
GT transitions whose strength are a small fraction of the sum rule the
direct relationship between  $\sigma(p, \ n)$ and $B(GT)$ values
fails to exist. Similar discrepancies have been observed \cite{Wat85}
for reactions on some
odd-A nuclei including $^{13}C$, $^{15}N$, $^{35}Cl$, and $^{39}K$ and
for charge-exchange induced by heavy ions \cite{Ber96}.  

\subsection{Coulomb Dissociation Method}

We now consider projectile breakup reactions $a+A \longrightarrow
b+x+A$ in which (a) the target $A$ remains in the ground state, 
(b) the nuclear contribution to the breakup is small, and
(c) the breakup is a first order process. Then
the (differential, or angle integrated) breakup cross section can be written
as $\sigma_C^{\pi\lambda} (\omega) = F^{\pi\lambda}(\omega) \ .\  
\sigma_\gamma^{\pi\lambda} (\omega) $, where $\omega$ is the energy
transferred from the relative motion to the breakup, and
$\sigma_\gamma^{\pi\lambda} (\omega)$ is the photo nuclear cross section
for the multipolarity  ${\pi\lambda}$ and photon energy $\omega$.
The function $F^{\pi\lambda}$ depends on $\omega$, the 
relative  motion energy, and nuclear charges and radii. They can be very 
accurately calculated \cite{Ber88}
for each multipolarity ${\pi\lambda}$. 

Time reversal allows one to deduce the radiative capture cross section
$b+x \longrightarrow a+\gamma$ from $\sigma_\gamma^{\pi\lambda} (\omega)$. Thus,
assuming that the conditions (a--c) apply, the Coulomb dissociation
can be used to deduce $\sigma (b+x \longrightarrow a+\gamma)$. This 
method was proposed in ref. \cite{BBR86}. 
It has been tested successfully in a 
number of reactions of interest for astrophysics. For example, the
$d(\alpha, \ \gamma ) ^6Li$, relevant for the evolution of the primordial
fireball
has been measured down to 100 keV by using the breakup of 156 MeV $^6Li$ ions
on several targets \cite{Kie91}. They have 
obtained the only 5 points in the energy interval 100--710 keV, previously
unaccessible by direct capture experiments. 
Another case was the $^{13}N(p, \ \gamma)^{14}O$ reaction which was studied
via the $^{14}O \longrightarrow ^{13}N + p$ Coulomb dissociation on heavy
targets \cite{Mot91}.      
The $^{13}N(p, \ \gamma)^{14}O$ reaction is very important for the
CNO cycle when its rate becomes faster than $^{13}N\beta$-decay. This leads to
a breakout of the CNO-cycle. The rate of this reaction is essentially determined
by the gamma width $\Gamma_\gamma$ of the 5.173 MeV $1^-$ resonance of $^{14}O$.
The value obtained with the Coulomb dissociation method (CDM) $\Gamma_\gamma =
(3.1 \pm 0.6)$ eV, compares well with the result from direct capture 
experiments. For a review of the results obtained with the CDM,
see ref. \cite{Bau94} and
references therein.
Worth quoting is the recent experiment  where the reaction $^7Be(p, \gamma)^8B$
has been studied with the CDM \cite{Mot94}. 
They have obtained an $S_{17}(0)$ value of
16.7 eV.b which is about 20\% smaller than the value commonly
used in solar model calculations \cite{Bah89}.
However, this CDM experiment yields a rather large $E2$ breakup
cross section which has to be subtracted from the total breakup
cross section in order to allow the determination the E1 radiative
capture cross, which dominates at low $^7Be-p$ energies. Work in
this direction is under way in several laboratories. For this
reaction, the dominance of
the Coulomb interaction in the dissociation process, and the validity of
first-order perturbation theory have been demonstrated theoretically
\cite{Ber95}.

In storage rings a series of reactions could be studied using
the CDM, depending on the luminosities: (a) $^4He(d, \gamma)^6Li$, $^6Li(p, \gamma)^7Be$,
$^6Li(\alpha, \ \gamma)^{10}B$, $^4He(t, \gamma)^7Li$, 
$^7Li(\alpha, \ \gamma)^{11}B$, $^{11}B(p,\ \gamma)^{12}C$,
$^9Be(p, \ \gamma)^{10}B$, and $^{10}B(p, \ \gamma)^{11}C$,
[primordial nucleosynthesis of Li, Be, and B-isotopes], (b)
$^{12}C(n, \ \gamma)^{13}C$, $^{14}C(n, \ \gamma)^{15}C$, and
$^{14}C(\alpha, \ \gamma)^{18}O$ [inhomogeneous Big Bang
nucleosynthesis], (c) $^{12}C(p, \ \gamma)^{13}N$,
$^{16}O(p, \ \gamma)^{17}F$, $^{13}N (p, \gamma)^{14}O$,
and $^{20}Ne(p, \ \gamma)^{21}Na$ [CNO-cycle], 
(d) $^{31}S(p, \ \gamma)^{32}Cl$ [rapid proton capture process], and
(e) $^{12}C(\alpha, \ \gamma)^{16}O$, $^{16}O(\alpha, \ \gamma)^{20}Ne$,
and $^{14}N(\alpha, \ \gamma)^{18}F$ [helium burning].   
The projectiles are either stable, or have a half-life larger than a few seconds.
Of course, one of the most interesting cases is the reaction $^{12}C(\alpha, \ \gamma)
^{16}O$, or the triple-alpha capture reaction $3\alpha \longrightarrow ^{12}C$, both
relevant for the fate of massive stars. However, due to the large energy ($\sim 7$ MeV)
required to disrupt $^{12}C$, or $^{16}O$, the nuclear contribution to
the breakup process is large. Nonetheless, Coulomb dissociation
might be an alternative if special kinematical conditions 
are chosen.


\begin{thebibliography}{9}
\bibitem{Sal54}  E.E. Salpeter, Aust. J. Phys. 7 (1954) 373
\bibitem{Ass87}  H.J. Assenbaum, K. Langanke and C. Rolfs, Z. Phys. A327 (1987) 461                 
\bibitem{Eng88}  S. Engstler {\it et al.}, Phys. Lett. B202 (1988) 179
\bibitem{Blu90}  G. Bl\"ugge and K. Langanke, Phys. Rev. C41 (1990) 1191; K. Langanke and D. Lukas, Ann. Phys. 1 (1992) 332
\bibitem{Rol95}  C. Rolfs and E. Somorjai, Nucl. Inst. Meth. B99 (1995) 297
\bibitem{Kaj95}  T. Kajino, Nucl. Phys. A588 (1995) 339c
\bibitem{Par90}  T. Paradellis {\it et al.}, Z. Phys. A337 (1990) 211
\bibitem{Boy92}  R.N. Boyd {\it et al.}, Phys. Rev. Lett. 68 (1992) 1283
\bibitem{Bah89}  J.N. Bahcall,``Neutrino Astrophysics", Cambridge University Press, 1989
\bibitem{Bar80}  F.C. Barker, Aust. J. Phys. {\bf 33} (1980) 177;
Phys. Rev. {\bf C37} (1988) 2930
\bibitem{Hoy53}  F. Hoyle, D.N. Dunbar, W.A. Wenzel and W. Whaling, Phys. Rev. 92 (1953) 1095
\bibitem{Woo85}  S.E. Woosley, Proceedings of the Accelerated Radioactive Beam Workshop,
eds. Buchmann and J.M. D'Auria (TRIUMF, Canada, 1985).
\bibitem{Buch93}  L. Buchmann {\it et al.}, Phys. Rev. Lett. 70 (1993) 726; Z. Zhao {\it et al.},
Phys. Rev. Lett. 70 (1993) 2066 
\bibitem{Ober96} H. Oberhummer, H. Herndl, T. Rauscher and H. Beer, University of Vienna, preprint 1996, to be published
\bibitem{Jun92}  M. Jung {\it et al.}, Phys. Rev. Lett. 69 (1992) 2164
\bibitem{Bos95}  F. Bosch {\it et al.}, GSI/Darmstadt Scientific Report, 1995
\bibitem{Bos96}  F. Bosch, Spokesperson, GSI/Darmstadt experiment
proposal E019, 1996
\bibitem{Neu95}  S. Neumaier {\it et al.}, Nucl. Phys. A583 (1995) 799
\bibitem{Kor96}  A.A. Korsheninnikov {\it et al.}, Proceedings of the
Int. Workshop on Physics of Unstable Nuclear Beams, (Serra Negra, Brazil, 1996),
eds. C.A. Bertulani, L.F Canto and M.S. Hussein, 
World Scientific, in press
\bibitem{Egel96} P. Egelhof and K.L. Kratz, spokepersons, GSI/Darmstadt proposal E023 
\bibitem{San82}  D.G. Sandler, S.E. Koonin and W.F. Fowler, Ap. J. 259
(1982) 908
\bibitem{Wie96}  M. Wiescher, private communication
\bibitem{Bau86}  G. Baur, Phys. Let. B178 (1986) 135
\bibitem{Muk90}  A.M. Mukhamedzhanov and N.K. Timofeyuk, JETP Lett. 51 (1990) 282
\bibitem{Wei96}  Weiping Liu {\it et al.}, Phys. Rev. Let. 77 (1996) 611
\bibitem{Tad87}  T.N. Tadeucci {\it et al.}, Nucl. Phys. A469 (1987) 125
\bibitem{Aus94}  S.M. Austin, N. Anantaraman and W.G. Love, Phys.
Rev. Lett. 73 (1994) 30
\bibitem{Wat85}  J.W. Watson et al., Phys. Rev. Lett. 55 (1985) 1369 
\bibitem{Ber96}  C.A. Bertulani and P. Lotti, to be published.
\bibitem{Ber88}  C. Bertulani and G. Baur, Phys. Reports 163 (1988) 299
\bibitem{BBR86}  G. Baur, C. Bertulani and H. Rebel, Nucl. Phys. A459 (1986) 
188
\bibitem{Kie91}  J. Kiener et al., Phys. Rev. C44 (1991) 2195
\bibitem{Mot91}  T. Motobayashi et al., Phys. Lett. B264 (1991) 259
\bibitem{Bau94}  G. Baur and H. Rebel, J. Phys. G20 (1994) 1
\bibitem{Mot94}  T. Motobayashi et al., Phys. Rev. Lett. 73 (1994) 2680
\bibitem{Ber95}  C. Bertulani, Nucl. Phys. A587 (1995) 318; Phys. Rev. C49 
(1994) 2688; Z. Phys. A, to be published


\end{thebibliography}
\end{document}